# High Performance Computing for Geospatial Applications: A Prospective View

Marc P. Armstrong
The University of Iowa
marc-armstrong@uiowa.edu

**Abstract**

The pace of improvement in the performance of conventional computer hardware has slowed significantly during the past decade, largely as a consequence of reaching the physical limits of manufacturing processes.  To offset this slowdown, new approaches to HPC are now undergoing rapid development. This chapter describes current work on the development of cutting-edge exascale computing systems that are intended to be in place in 2021 and then turns to address several other important developments in HPC, some of which are only in the early stage of development. Domain-specific heterogeneous processing approaches use hardware that is tailored to specific problem types. Neuromorphic systems are designed to mimic brain function and are well suited to machine learning. And then there is quantum computing, which is the subject of some controversy despite the enormous funding initiatives that are in place to ensure that systems continue to scale-up from current small demonstration systems.

**1.0 Introduction**

Rapid gains in computing performance were sustained for several decades by several interacting forces, that, taken together, became widely known as Moore's Law (Moore, 1965).  The basic idea is simple: transistor density would double approximately every two years.  This increase in density yielded continual improvements in processor performance even though it was long-known that the model could not be sustained indefinitely (Stone and Cocke, 1991).  And this has come to pass. Part of the reason for this performance fall-off is related to Dennard Scaling, which correlates transistor density with power consumption per transistor: for many years as density increased, power consumption per unit area remained near-constant.  As described by Hennessy and Patterson (2019: 53) this relationship held up well until around 2007, when it began to fail, and fail big-time. This failure is a reason why clock frequencies have not continued to increase rapidly during the last decade. Instead, computer architects turned to the use of multiple cores to increase performance.  This multicore path, however, raised other issues such as inefficiencies that arise as a consequence of ineffective thread exploitation, as well as failures in speculative branch prediction, which, in turn, causes wasted processing effort, excessive power consumption.  Indeed, the use of increasing numbers of cores push chips past their thermal design limits and require selective power-gating approaches to prevent thermal failures (Perricone et al. 2018: 60).  This has led to the problem of "dark silicon" in which cores only operate part-time.  These problems paint a dire performance picture, but Hennessy and Patterson show a path forward, one that will require some re-thinking about programming and the use of novel architectural arrangements. This chapter describes the near-future state of exascale computing and then turns to a discussion of other ways to wrest improved performance out of both existing and experimental technologies including heterogeneous processing as well as neuromorphic and quantum computing.

**2.0 The pursuit of exascale systems**

There is now a race to construct an exascale system that will be able to provide a quintillion floating point calculations per second (Normile, 2018). What is that? 1,000,000,000,000,000,000/second or a thousand petaflops. Exascale systems are viewed as strategic initiatives by several national governments that have made significant investments to support their development. While these investments can be considered to be part of ongoing support for scientific research, other rationales include national prestige and military defense (*e.g.*, simulating nuclear explosions). This is an area of very rapid change and international intrigue, with major initiatives being pursued by the US, China and Japan.

In the US, the Aurora Project is being funded by the Department of Energy to the tune of $500 million. Located at Argonne National Laboratory, Aurora will use a Cray Shasta architecture with Intel nodes built using a new generation of Xeon Scalable processors and a new $X^e$ architecture for GPUs. Cray Slingshot communications technology provides Aurora with very high bandwidth communication: 25.6 Tb/s per switch, from 64 200 Gbs ports[1].

China had topped the "Top500" list of supercomputers for several years prior to being displaced by the US in 2018. The Chinese government is now funding three projects, at levels generally reported to be in the "several billion dollar" range, to help reach the goal of exascale performance levels[2]. Each project has constructed prototype systems as a proof-of-concept. It is notable that all Chinese high performance implementations since 2015 use domestically-sourced processors because of technology trade barriers put in place by the US government. This has given a substantial boost to the Chinese chip manufacturing industry.

1. The National Research Center of Parallel Computer (NRCPC) prototype uses Sunway SW26010 processors that are configured into 512 two-processor nodes connected by a locally-developed network. Though each processor has four quadrants with its own management core and 64 compute cores, to reach exascale performance, the system will need improved processors and a lot of them.
2. The Sugon system is heterogeneous and uses a Hygon clone of an AMD x86 processor. The system is constructed with nodes consisting of two Hygon processors and two data cache unit (DCU) prefetch accelerators that are connected by a 6D torus network. It was recently reported that Sugon was going to demonstrate a world-class machine (using an x86 processor and 4 AMD GPUs per blade) at a recent supercomputing conference, but held back in order not to shed light on its prowess. That tactic did not work and Sugon has recently been denied access to US technologies by the US Department of Commerce[3] after being placed on the so-called "entity list".
3. The National University of Defense Technology (NUDT) is also building a heterogeneous system comprised of CPUs and digital signal processor chips that each have 218 cores. The system is configured using a series of "blades" containing eight CPUs and eight DSPs. Each blade will provide 96 teraflops and will be housed in a cabinet holding 218 blades; with 100 of these cabinets installed, a total of 1.29 peak exaflops will be possible.

---

[1] www.cray.com/sites/default/files/Slingshot-The-Interconnect-for-the-Exascale-Era.pdf

[2] https://www.nextplatform.com/2019/05/02/china-fleshes-out-exascale-design-for-tianhe-3/

Not to be outdone, Japan is in the process of designing and implementing a new system with the name of Fugaku (a nickname for Mt Fuji) that will be completed in 2021. Funded by the Japanese government at a level of approximately $1 billion through their RIKEN institute, Fugaku will be constructed using Fujitsu A64FX CPUs. Each chip has a total of 48 compute cores: four core memory group modules with 12 cores and a controller core, L2 cache and memory controller and other architectural enhancements[4]. These chips will be connected using Tofu (torus fusion) with six axes: X, Y, Z that vary globally according to the size of the system configuration while A, B, C are local (2x3x2) toroidal connections.

The search for processing supremacy will undoubtedly continue beyond the exascale milestone. It should also be noted that an exaflop level of performance is not meaningful without a software environment that enables programmers to harness the full potential of the hardware.

**3.0 Coding practices**

One well-known way to improve performance is to re-write software in more efficient code. Hennessy and Patterson (2019:56) cite the example of a particular matrix multiplication code first written in Python, a scripting language that uses inefficient dynamic typing and storage management. The code was then re-written in C to improve cache hits and to exploit the use of multiple cores and other hardware extensions. This re-casting led to substantially increased performance. It should be noted, however, that improvements of this type are well-known. For example, early geospatial software was written partly in assembler to boost the performance of core computational tasks and MacDougall (1984: 137) also reports results for interpolation software that was written first in Basic and then in C. For the largest problem size reported, this language switch (from interpreted to compiled) reduced execution time from 747.4 to 5.9 minutes. Though such changes require significant programming intervention, it is clear that substantial performance improvements can be realized.

**4.0 Domain-specific computing and heterogeneous processing**

Hennessy and Patterson (2019) also point to a more architecturally-centric approach to performance gains: domain-specific architectures. Their position represents a significant expansion of concepts that have a long history. Examples include math co-processors for early x86 chips, and graphics processing units (GPUs) that have their roots in the 1980s. Though floating-point units and graphics processing have now been incorporated into the functionality of commodity CPUs, a key aspect of domain specific computing is the identification of a particular type of architecture that is most germane to a specific application or problem class. Research is now being conducted to introduce new types of specialized hardware that can be applied to specific types of problems, and work is proceeding apace on the development of software tools that can be used to harness disparate architectures. Zahran (2019) provides a comprehensive look at the current state of the art in both the hardware and software used to implement these heterogeneous systems.

---

[3] https://www.hpcwire.com/2019/06/26/sugon-placed-on-us-entity-list-after-strong-showing-at-isc/
[4] https://www.fujitsu.com/global/Images/post-k-supercomputer-development.pdf

Arguments were advanced decades ago in support of the use of domain-specific architectures and heterogeneous processing (Siegel, Armstrong and Watson, 1992; Freund and Siegel, 1993). Siegel and associates describe a vision in which different code components are matched to architectures that are best suited to their execution. As Moore's Law has faltered, interest has been renewed in a kind of neo-heterogeneity that has been further expanded to consider energy efficiency. And the view of heterogeneity now includes a number of new types of processing devices.

- Field-programmable gate arrays (FPGA), are a type of integrated circuit that can be re-configured after it is manufactured (soft hardware). Intel, for example, has developed a commercial FPGA product called Stratix 10. It is also important to note that FPGAs normally consume far less power than CPUs or GPUs (Alonso, 2018: 2).
- Jouppi et al. (2018) describe a unique architecture that has developed to support the execution of deep neural networks (DNN). A tensor processing unit (TPU) is optimized to perform matrix multiplications and outperform CPUs and GPUs in both number of operations and energy efficiency. Part of this efficiency is achieved through the use of a single, large two dimensional multiply unit, the removal of features normally required by CPUs, and the use of eight-bit integers since DNN applications do not require high precision computation, but do require very large numbers of training cycles. In effect, a TPU is optimized to perform lots of inexpensive, low-precision matrix operations.
- Intel is also developing an approach to supporting deep learning with an application specific integrated circuit (ASIC) product known as the Nervana Engine[5]. Having also rejected GPU approaches, Nervana has settled on 16 bit precision and each ASIC contains 6 links that enable it to be connected in a torus configuration to improve communication. A key aspect of training deep neural networks is data movement. Given the need for large numbers of training examples, to avoid data starvation the Nervana Engine provides 32 GB of storage with 8 terabits/second of bandwidth.
- Corden (2019) describes the development of vectorization instructions in Intel processors. In the example provided (a double precision floating-point loop), a comparison is made among SSE (Stream SIMD Extensions), AVX (Advanced Vector Extensions) and the newest AVX-512 which increase the vector width to 512 bits. In processing the same loop, scalar mode produces one result, SSE produces two results, AVX produces four results, and AVX-512 produces eight results. This represents a substantial advance in throughput for those applications that can exploit this architecture.

Figure 1 provides a simplified view of a collection of architectures that could comprise a heterogeneous system. Li, Zhao and Cheng (2016) describe a much more tightly integrated approach to linking CPU and

---

[5] https://www.intel.ai/nervana-engine-delivers-deep-learning-at-ludicrous-speed/#gs.k0p7wf

FPGA modes. Their CPU+[GPU, DSP, ASIC, FPGA] approach embeds heterogeneity either on chip or coupled with a high speed bus (Figure 2).

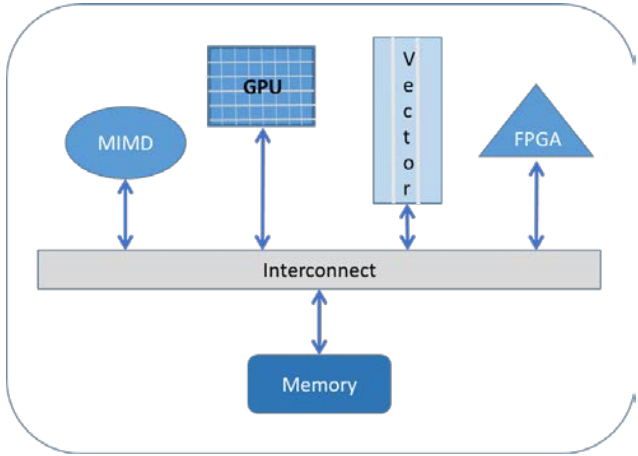

Figure 1. A stylized view of a collection of heterogeneous processors accessing shared memory.

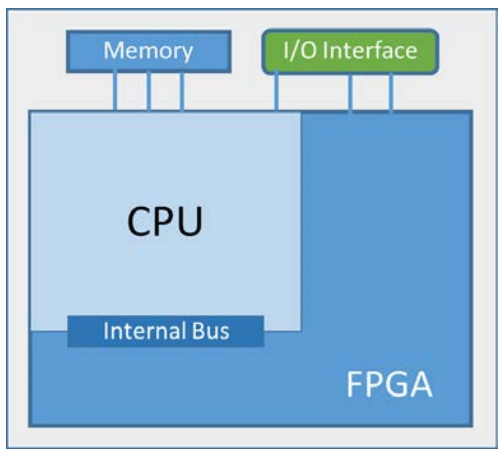

Figure 2. Schematic architecture of an integrated heterogeneous processor incorporating a CPU and FPGA. (Figure based on Li, Zhao and Cheng, 2016)

While programming is hard, parallel programming is harder and writing efficient programs for heterogeneous environments is exceedingly difficult. Zahran (2016:9) points to several measures of success against which heterogeneous software can be assessed:

- Performance (as measured by speedup)
- Scalability (number of homogeneous cores and across different architectures)
- Reliability (as determined by graceful degradation with core reductions and transient faults)
- Portability (across different architectures)

Siegel, Dietz and Antonio (1996:237) describe several factors that must be considered when writing heterogeneous programs including matching a sub task to a specific machine architecture, the time to move data that is analyzed on different machines, and the overhead incurred in stopping a task and restarting it on a different machine.  This entails problem decomposition into subtasks, assigning subtasks to machine types, coding subtasks for specific target machines, and scheduling the execution of subtasks.  Because of this complexity, researchers have searched for ways to hide it by using abstraction. One view advanced by Singh (2011) is to develop language agnostic multi-target programming environments that are distantly related to the idea of device independence in computer graphics. In the abstract example presented, a data parallel program would be compiled and executed using different target architectures that would vary according to runtime performance and energy reduction requirements (Figure 3).

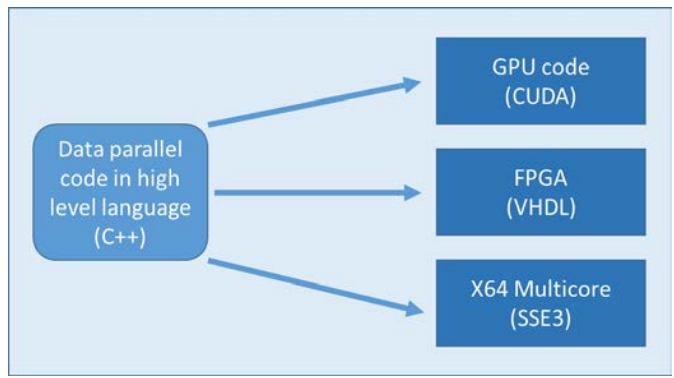

Figure 3. Compilation to multiple target architectures. CUDA is the Compute Unified Device Architecture developed by Nvidia for GPU devices; VHDL is Very High Speed Integrated Circuit Hardware Description Language which enables flexible configurations of FPGAs and other devices; SSE3 is Streaming SIMD Extensions, an evolving collection of low-level instructions that was developed by Intel. (Source: Based on Singh, 2011:3)

Reichenbach *et al*. (2018) describe an application that uses a library (LibHSA) that enables programmers to insert custom accelerators (e.g., to perform image processing with Sobel and Laplacian edge detectors) to accomplish tasks in heterogeneous environments. The library is compliant with an emerging standard model for data parallel processors and accelerators supported by the non-profit HSA Foundation that is working toward the goal of making it easier to program heterogeneous computing devices. The approach aims at reducing complexity and works with standard programming languages (e.g., Python) while abstracting low-level details to simplify the user view.  It is worth noting that the HSA Foundation was founded by key hardware players such as AMD, ARM, Samsung, Texas Instruments and others (see: hasfoundation.com).

The argument for domain specific computing advanced by Hennessey and Patterson is worth considering in a geospatial context. While it is unlikely that specific electronic devices will be developed specifically to support spatial data analysis, *a la* the TPU, there are good reasons for spatial middleware to closely align the specific characteristics of geospatial algorithms to particular types of hardware environments. For example, some geospatial problems lend themselves to SIMD architectures, while others are more naturally suited to MIMD architectures and then, of course there are data distribution issues that affect efficiency (*e.g.*, Armstrong and Densham, 1992; Marciano and Armstrong, 1997; Cramer and Armstrong, 1999).

Densham and Armstrong (1994) sketched out a sequence of steps that would be required to run a vertex-substitution location-allocation model in a heterogeneous environment, though they did not advance it into implementation. Table 1 is an updated version of their sequence that includes new architectures and algorithms.

Table 1. Sequence of steps required to construct a location-allocation model for decision support in a heterogeneous environment.

| Sequence | Operation Type | Time Complexity | Architecture |
|---|---|---|---|
| 1 | Shortest Path | $O(n^4)$ or less | GPU |
| 2 | Create Strings | Sort: $O(n^2)$ or $(n \log n)$ | MIMD Column Sort |
| 3 | Hillsman Edit | Modify: $O(n)$ | MIMD Edit Each Element |
| 4 | Vertex Substitution | Substitute: $O(p*(n-p))$ | GPU |
| 5 | Visualization | Variable | GPU |

Note: Step 1 is due to Arlinghaus *et al*. (1990); step 4 is due to Lim and Ma (2013). Since vertex substitution is a heuristic method, it is often run numerous times to try to escape local optima.

**5.0 A few words on the changing nature of storage**

At the present time, storage technology is undergoing a phase-shift from spinning disks to solid state drives (SSD) and this shift has become quite pronounced at the high end where custom made SSDs are now widely used in data centers. This technology is under considerable pressure given the pressing requirements caused by the multiple dimensions of big data (*e.g.*, volume, velocity, and variety). It turns out, however, that for backup storage, tape continues to be an excellent choice in terms of density, capacity, durability and energy efficiency and there are also fruitful avenues that can be exploited to improve the performance of this medium (Greengard, 2019).

SSDs are designed with flash storage and a firmware controller that is responsible for managing the memory resource by doing low level tasks such as garbage collection and maintaining rough equality of writing to all storage areas to prolong the life of the device. Do, Sengupta and Swanson (2019) describe significant disruptive trends in SSD design. There is now a move away from the current bare bones approach that incorporates compute and software flexibility in the SSD, to a model in which each storage device has a greater number of cores and increased clock speeds, as well as increased flexibility provided by a more general purpose embedded operating system that will support upgrades. By moving computation closer to the data, akin to edge computing, there are corresponding benefits that accrue in many applications, including improved bandwidth and deceased latency, both of which are extremely important to fast data applications in the IoT.

**6.0 Neuromorphic Computing**

Neuromorphic computer systems are designed to emulate brain functions using spiking neural networks and are said to be superior to traditional DNNs because they are somewhat immune to a problem called catastrophic interference; when novel inputs are presented to a neuromorphic system, they can flexibly adapt to them and do not require complete re-training, as do DNNs. While this approach is in early stages of development and has been downplayed by some AI researchers, it is also clear that there are very large investments being made in the technology. For example, Intel's product is called Loihi[6] and it contains 128 neuromorphic cores. Intel has announced plans to construct a scalable system with 100 million neurons by 2020 (Moore, 2019). It is also notable that the Loihi chip set consumes far less power than a GPU or CPU.

IBM's TrueNorth program is a strong competitor.  Developed initially by simulating neuromorphic systems on conventional supercomputers such as BlueGene/Q, the simulation project eventually was instantiated in hardware.  The current generation is configured in a hierarchical arrangement with 16 TrueNorth chips (1 million neurons each) placed in a 4 x 4 grid connected by a custom I/O interface; four of these systems are used to create the NS16e-4 system with 64 million neurons and 16 billion synapses (DeBole *et al.*, 2019).  What is truly impressive is that this assemblage is extremely energy efficient, operating at approximately the same level as an incandescent light bulb (70 W).  Even though a neuromorphic supercomputer and a data center are very different, there is a sharp contrast between 70 W and the megawatt power consumption of current data centers.  Facebook, for example, has invested in a 138 megawatt wind farm to offset electrical grid power consumption at its Altoona, IA data center[7].

Application domains for neuromorphic computing include image recognition and feature extraction, as well as robotic and vehicle path planning.  The TrueNorth NS16e-4 system, for example, is being used to detect and classify objects in high-definition aerial video at greater than 100 frames per second (DeBole *et al.*, 2019: 25).

**7.0 Technology Jumps and the Quantum Quandary**

Though Hennessey and Patterson have painted a relatively grim picture about the future of high performance computing, other researchers have used a somewhat sunnier pallete. Denning and Lewis (2017) for example, conclude that Moore's Law level of performance can be sustained as a consequence of assessing performance not at the chip level, but by expanding the assessment to include entire systems and communities. And perhaps more importantly, they suggest several ways in which jumps can occur to escape the inherent limiting embrace of Moore's CMOS-based law and include alternative technologies with far greater promise of high performance. They are not alone in holding such views. The IEEE, for example, has a well-maintained web site[8] concerned with the design and application of experimental approaches to computing. Among the most widely discussed technologies are:

- Three-dimensional fabrication and packaging to escape planar form factors
- Using spintronics for data representation and switch construction
- DNA and biological computing
- Quantum computing

---

[6] https://en.wikichip.org/wiki/intel/loihi

[7] https://www.datacenterknowledge.com/archives/2014/11/14/facebook-launches-iowa-data-center-with-entirely-new-network-architecture

[8] https://rebootingcomputing.ieee.org/

The last element in this list, quantum computing, has the potential to be a "game changer" by exponentially improving computational performance. Such performance increases have several important applications in the geospatial domain, chiefly in the areas of combinatorial spatial optimization and machine learning. How might this work? Traditional computer systems, at their root, use bits (0 or 1) and collections of bits (*e.g.* bytes) to represent values. A quantum computer uses a radically different approach, using quantum bits or *qubits* that can assume either values of 0 or 1, as well as both at the same time using the quantum feature of superpositioning. This means that qubits can inhabit all possible states that can be assumed by a traditional computer, thus enabling work in an exponentially larger problem space with vastly improved performance (NAS, 2019:2). As a consequence, there is considerable interest in advancing quantum computing capabilities, and a kind of arms race has ensued in part because a fully functioning quantum computer could rapidly (in polynomial time) factor integers and therefore decrypt public key encoded messages (*e.g.*, RSA). In support of these activities, the National Quantum Initiative Act[9] was signed into law in 2018. The Act enables the Department of Energy to provide $1.25 billion to build interest in the business community, as well as to support the establishment of quantum computing national research centers. Similar types of funding initiatives are taking place in China, Japan and Europe.

It is important to recognize, however, that there are many large barriers to overcome before a fully functional, large quantum computer is implemented. Some of these may take a decade or more to address, as described in a NAS Consensus Study Report (2019: 2-11).

- Unlike traditional binary computers that have large capacities for noise rejection, qubits can be any combination of one and zero, and as a consequence, quantum gates cannot reject even small amounts of noise, since it can be confused with an actual state.
- Though quantum error correction can be used to reduce the effects of noise and yield error-corrected results, the amount of processing overhead required to achieve this goal is prohibitive.
- The conversion of "normal" digital data into a quantum state is time consuming and would dominate processing time requirements in an Amdahl's Law-like fashion; this is particularly problematic for geospatial big data applications.
- New algorithms and a new software stack will be needed.
- Debugging is exceedingly difficult. Intermediate states are not determinable since any measurement of a quantum state halts execution; when a register is read, superposition collapses.

Because of these problems, a number of researchers are beginning to question the viability of the entire quantum enterprise (see, for example, Edwards, 2019 and Vardi, 2019). This skepticism is reinforced by the absence of a so-called virtuous cycle that has characterized developments in traditional integrated circuits during the past several decades: a voracious market generated large profits that were reinvested in research and development that then led to new products that created new revenue streams. In the absence of such a cycle, it could be many years before a viable quantum product is created. It should be noted, however, that the Quantum Initiative Act is intended to jump-start such a cycle.

---

[9] https://www.congress.gov/115/bills/hr6227/BILLS-115hr6227enr.xml

Despite these concerns, researchers are moving forward with work using the relatively small quantum devices that are now available. Though these devices are not able to handle practical applications, quantum computers will need to scale in order to be successful. One approach decomposes problems into smaller parts using classical computing systems and the sub-components are allocated to quantum processing. This approach incurs an overhead processing penalty and as described by Shaydulin *et al*. (2019: 20) two $n$-qubit computers would be less powerful and efficient than one $2n$-qubit processor. However, if the quantum part of a complex problem can be effectively exploited, the penalty is minor, a kind of inverse Amdahl's Law, as it were.  DeBenedictis, Humble and Gargini (2018) outline a further set of requirements that need to be satisfied to reach the objective of quantum scalability.  They begin by describing the slow original progress made by metal-oxide semiconductors and suggest that a similar path be developed for quantum systems, citing the work of DiVincenzo (2000) who lists several criteria, including:

- Physical scalability with well-defined qubits
- A generalized quantum gate structure
- Decoherence times that are substantially greater than gate operation

 The first criterion is an echo of a Moore's Law-like concept. The second is a paean to the march of technology, while the third may prove to be an intractable difficulty, though DeBenedictis, Humble and Gargani (2018) liken the problem to the quest to reduce defects in integrated circuits.  Nevertheless, it is also important to recall that now unforeseen innovations can fundamentally alter the trajectory of technological developments.

**8.0 Summary and Conclusion**

This chapter has provided an overview of several responses to the end of Dennard scaling as applied to the manufacture of conventional CMOS components.  These responses become important when considering the substantial computational complexity of many geospatial methods, particularly when they are applied to increasingly large databases.  For example, increases in the volume and velocity of data streamed from internet-connected devices will accelerate as the Internet of Things continues to infiltrate into routine daily activities.  Though these new approaches to computing hold promise, they also present interesting challenges that will need to be addressed, in many cases, by inventing new algorithms and revising software stacks.

The next generation of exascale systems will exploit conventional advances, but such systems are difficult to sustain and require enormous quantities of power to operate.  And software that can efficiently exploit thousands of cores in a mixed CPU/GPU architecture is difficult to create.  Instead of continuing along the same path, new approaches are being developed to match problems to specialized processing elements and to develop new classes of hardware that do not follow traditional architectural models.

Advances are being made in the development of new heterogeneous systems that use conventional CPUs, but also deploy other architectures, such as GPUs, TPUs and FPGAs. Related progress is also being made on the development of programming models that are designed to enable better matches between architectures and model components, though at the present time, programmer intervention is required.

Neuromorphic computing, which attempts to emulate cognitive functioning, is viewed by some with skepticism, though it is also clear that enormous investments in the development of hardware are being made by major component manufacturers. A key aspect of the approach is that it is designed to escape the brittleness of current deep neural networks that work well for a limited range of problems, but fail when they are presented with novel inputs. This fragility problem is described by Lewis and Denning (2018) and amplified by a collection of letters in response (see Communications of the Association for Computing Machinery, 08/2019: 9).

Quantum computing remains a "known unknown" at the present. There is much promise, as well as tremendous hype, and quantum scaling seems to be accepted as inevitable by some. One article in the non-technical press sums it up: "Google believes it will reach 'quantum supremacy' – a stunt-like demonstration of a machine's superiority over a traditional computer – in the very near term. Chinese scientists say they're on a similar timeline." (Hackett, 2019: 164). This is far from the pessimistic tone promulgated by the recent NAS report. But Hackett also quotes a researcher who advocates for patience, indicating that Sputnik was launched in 1957, but Neil Armstrong didn't leap onto the Moon until 1969. And while the transistor was invented in 1947, the first integrated circuit wasn't produced until 1958. So, it seems as if time will tell with quantum computing.